\documentclass{aa}  
\usepackage{hyperref}
\hypersetup{colorlinks, citecolor=blue, linkcolor=blue, urlcolor=blue}
\usepackage{graphicx}
\usepackage{txfonts}

\begin{document}

   \title{Cosmic rate of type IIn supernovae \\ and its evolution with redshift}

   \author{C. Cold
          \inst{1}
          \and
          J. Hjorth\inst{1}
          }

   \institute{DARK, Niels Bohr Institute, University of Copenhagen, Jagtvej 128, 2200 Copenhagen N, Denmark\\
             }

   \date{Received month day, year; accepted month day, year}

  \abstract  
   {Type IIn supernovae potentially constitute a large fraction of the gravitationally lensed supernovae predicted to be found with upcoming facilities. However, the local rate is used for these estimates, which is assumed to be independent of properties such as the host galaxy mass. Some studies hint that a host galaxy mass bias may exist for IIn supernovae.   
   }
   {This paper aims to provide an updated local IIn supernova-to-core-collapse ratio based on data from the Palomar Transient Factory (PTF) and the Zwicky Transient Facility (ZTF) Bright Transient Survey (BTS). Furthermore, the goal is to investigate the dependency of the IIn supernova peak magnitude on the host galaxy mass and the consequences of a possible host galaxy mass preference on the volumetric rate of type IIn supernovae.
   }
   {We constructed approximately volume-limited subsamples to determine the local IIn supernova-to-core-collapse ratio. We investigated the absolute peak magnitude of a subsample of type IIn and superluminous II or IIn supernovae exploring how this relates to the $i$-band magnitude of the host galaxies (as a proxy for stellar mass). We presented a method to quantify the effect of a potential preference for low-mass host galaxies utilizing the \textsc{UniverseMachine} algorithm.
   }
   {The IIn supernova-to-core-collapse ratios for PTF and BTS are $0.046 \pm 0.013$ and $0.048 \pm 0.011$, respectively, which results in a ratio of $0.047 \pm 0.009$, which is consistent with the ratio of 0.05 currently used to estimate the number of gravitationally lensed IIn supernovae. We report fainter host galaxy median absolute magnitudes for type IIn brighter than $-20.5$ mag with a 3 $\sigma$ significance. If the IIn supernova-to-core-collapse ratio were described by the power law model $IIn/CC = 0.15 \cdot \log(M/M_{\odot})^{-0.05}$, we would expect a slightly elevated volumetric rate for redshifts beyond 3.2. }


   \maketitle
%

\section{Introduction}
Type IIn supernovae (SNe IIn) exhibit narrow hydrogen emission lines in their spectra \citep{1990MNRAS.244..269S}. The distinct features of this SN class arise from the slow-moving and dense circumstellar material (CSM) ejected by the star prior to explosion. This means that the SN IIn subtype is very diverse, as these SNe can emerge whenever CSM indications are present in their spectra, whether it is early or late in the lifespan of the SN, or whatever lies beneath the veil of the CSM \citep{2017hsn..book..403S}. Narrow hydrogen features may also arise from flash ionization of local CSM following shock breakout. However, such emission lines disappear shortly after peak magnitude to reveal the underlying SN type \citep{2017NatPh..13..510Y, 2021ApJ...912...46B,  2022ApJ...924...15J, 2022ApJ...926...20T}. Nevertheless, flash ionization in combination with the complexity of the CSM structure can complicate the classification of SNe IIn \citep{2021MNRAS.506.4715R}.

Luminous blue variables, extreme red super giants, and yellow hyper giants have all been proposed as progenitors due to their recurring violent mass-loss episodes. The intervals of mass loss can range from a period of months to thousands of years, which is necessary to produce the amount of CSM required to make SNe IIn \citep{2017hsn..book..403S}. 

Brighter and more long-lived than other SN types, super-luminous supernovae (SLSNe) are recognized as their own class of SNe \citep{2018SSRv..214...59M, 2012Sci...337..927G}. These very luminous objects differ from other SNe by their optical absolute magnitudes of around $-21$ or less, although SLSNe have been classified at around $-19$ mag at peak for the faintest objects \citep{2018SSRv..214...59M, 2019MNRAS.487.2215A}. SLSNe, as the classic SN classes, are also further categorized into subtypes. The super-luminous counterpart to the SNe IIn, SLSNe-IIn, also feature narrow emission lines of the hydrogen Balmer series similar to the regular SNe IIn, and these constitute a significant percentage of all hydrogen-rich SLSNe \citep{2019ARA&A..57..305G}. This is indicative of CSM interaction partly powering very bright transients. It is not yet clear whether SLSNe-IIn and SNe IIn are two distinctive populations or if they form a continuum in luminosity. However, in this work, we considered the SLSNe-IIn as the brightest SNe IIn. 

Models indicate that SNe IIn along with SNe Ia will dominate the observed rates of lensed supernovae. Estimates from the upcoming Legacy Survey of Time and Space (LSST) with the Vera C. Rubin Observatory \citep{2019MNRAS.487.3342W, 2019ApJS..243....6G} predict on the order of 100 SNe IIn per year to be gravitationally lensed. For the \textit{Roman Space Telescope}, the gravitationally lensed SNe predictions are comparable \citep{2021ApJ...908..190P}. 

   \begin{figure*}[h!]
   \centering
   \includegraphics[width=\textwidth]{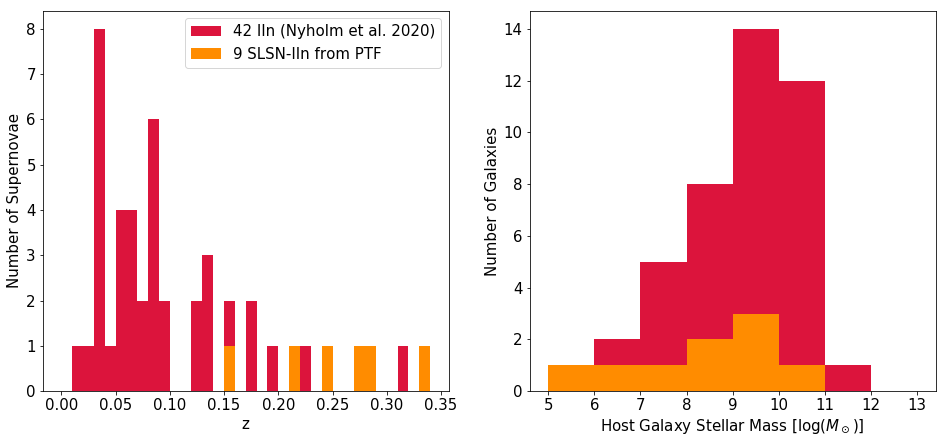}
   \caption{Overview of redshift and host galaxy mass distributions. {\it Left panel}: Redshift distribution of the SNe IIn in \cite{2020A&A...637A..73N} and SLSN-IIn from PTF. Both are subsamples of the full PTF SNe IIn and SLSNe-IIn samples. The redshifts can be found in \cite{2021ApJS..255...29S}. {\it Right panel}: Distribution of the host galaxy masses corresponding to the same SNe IIn from \cite{2020A&A...637A..73N} and a subsample of SLSNe-IIn from the PTF. These subsample distributions are consistent with the full sample distributions of SNe IIn and SLSNe-IIn from \cite{2021ApJS..255...29S}. }
              \label{nyholmdist}%
    \end{figure*}
    
\begin{figure*}[h!]
   \centering
   \includegraphics[width=\textwidth]{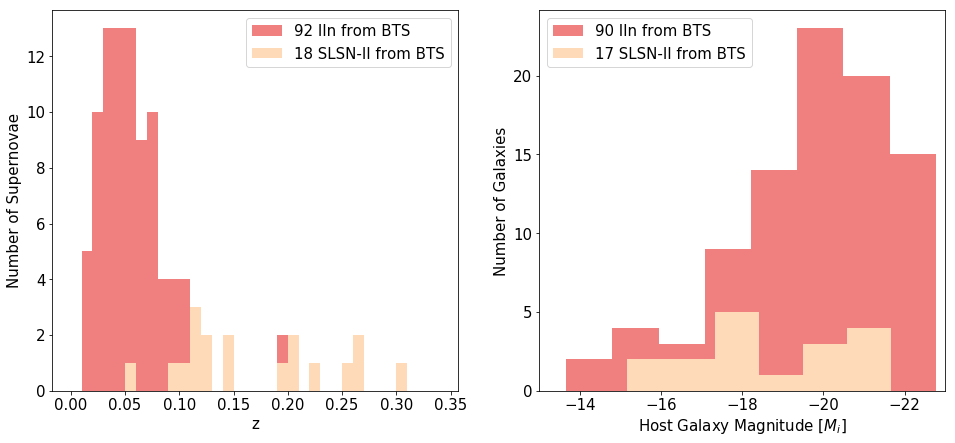}
   \caption{Overview of redshift and host galaxy magnitude distributions. {\it Left panel}: Redshift distribution of the SNe IIn and SLSN-II from BTS. {\it Right panel}: Distribution of the host galaxy $i$-band absolute magnitudes. For two SNe IIn and 1 SLSN-II, the $i$-band magnitudes were not available. The $i$-band magnitude is used as a proxy for the host stellar mass.}
              \label{btsdist}%
    \end{figure*}

The lensed SNe IIn predictions are based on the observed local rate of SNe IIn. Data from the Lick Observatory Supernova Search (LOSS) yielded an SNe-IIn-to-CC ratio of $8.8 \% ^{+3.3\%}_{-2.9\%}$ SNe IIn out of all CC SNe \citep{2011MNRAS.412.1441L, 2011MNRAS.412.1522S}. Several of these SNe IIn were subsequently identified as SN Impostors such that the IIn rate from LOSS is now considered to be around 5\% \citep{2017ApJ...837..121G}. Other examples of local rate measurements include studies by \cite{2009MNRAS.395.1409S}, who present a volume-limited (28 Mpc) sample compiled from all local SNe with a named host galaxy discovered over a ten-year period. This sample includes 3.8\% SNe IIn out of all CC SNe in the sample. \cite{2013MNRAS.436..774E} updated the study done by \cite{2009MNRAS.395.1409S}, searching for SNe discovered over a 14-year period, yielding $2.4 \pm 1.4\%$.

The existing rate estimates of SNe IIn assume that the local fraction of IIn to all other CC SNe does not evolve with redshift and so is not affected by large-scale changes in compositions or characteristics of galaxies and stellar populations over time. Several studies show a bias toward less massive host galaxies for SLSNe in general \citep[e.g.,][]{2015MNRAS.449..917L, 2016MNRAS.458...84A, 2018MNRAS.473.1258S, 2021MNRAS.503.3931T}, and this dearth of massive hosts suggests some dependence on the characteristics of the environment. In a study by \cite{2017ApJ...837..121G}, based on data from LOSS, SNe IIn seem to be more common in less massive galaxies. However, other studies do not come to the same conclusion \citep[e.g.,][]{2012ApJ...759..107K}. The Palomar Transient Factory (PTF) CC SN sample presented in \cite{2021ApJS..255...29S} also does not reveal a bias for SNe IIn toward low-mass host galaxies, and is inconclusive regarding the host galaxy mass preference of SLSNe-IIn.

The Zwicky Transient Facility (ZTF) Bright Transient Survey (BTS) \citep{2020ApJ...895...32F, 2020ApJ...904...35P} will be part of the analysis in this paper in addition to the PTF CC sample \citep{2021ApJS..255...29S}. The paper is structured as follows. In Section~\ref{section_data}, we briefly introduced the data used for the analysis. In Section~\ref{section_inferred_fractions}, we created an approximately volume-limited sample from the PTF and BTS data, followed by a presentation of an updated SNe-IIn-to-CC ratio (SNe IIn fraction) for both samples. In Section~\ref{section_brightness}, we compare the absolute magnitude and the $i$-band magnitude of the host galaxies of a subsample of SNe IIn, SLSNe-IIn, and SLSN-II from PTF and BTS. In Section~\ref{section_consequences}, we presented a generic method for inferring the rate and its evolution with redshift and studying the consequences of a possible mass-biased SNe IIn rate, before the discussion in Section~\ref{section_discussion} and conclusions in Section~\ref{section_conclusions}.

    
\section{Data}
\label{section_data}
The PTF CC SN sample from \cite{2021ApJS..255...29S} and the ZTF Bright Transient Survey \citep{2020ApJ...895...32F, 2020ApJ...904...35P} constitute the basis of the analysis presented in this paper. The PTF was a deep, wide-field survey followed by the intermediate PTF (iPTF) survey. The PTF CC sample contains 888 objects, of which 111 are SNe IIn and 16 SLSNe-IIn. Redshifts and host galaxy stellar masses are available for all objects in the sample. For later analysis, we will use the host galaxy $i$-band (either Pan-STARRS1 (PS1) or Sloan Digital Sky Survey (SDSS) $i$-band) absolute magnitude. However, two SNe IIn and one SLSN-IIn have no reported host $i$-band magnitude. The sample contains only host photometry, and so we used a subsample of IIn and SLSNe-IIn where the peak absolute magnitude is available. The subsample of 42 SNe IIn with available peak magnitudes is described in \cite{2020A&A...637A..73N}. These SNe were chosen based on the amount of available light-curve data to allow an analysis of both the rise times and decline rates of the SNe IIn, all with at least one available low-resolution spectrum. The redshifts in \cite{2020A&A...637A..73N} differ slightly from the redshifts in \cite{2021ApJS..255...29S}, which are the galaxy redshifts estimated by one of four possible methods: taken from SDSS, taken from the NASA Extragalactic Database, estimated from galaxy lines in the spectra or estimated from SN-template matching \citep{2021ApJS..255...29S}. In \cite{2020A&A...637A..73N}, the redshifts are estimated from the $\rm{H\alpha}$ emission lines in the SNe spectra. In this work, we used the data published in \cite{2021ApJS..255...29S} as well as SLSNe-IIn peak magnitudes (Leloudas, priv. comm.). The redshift and host galaxy stellar mass distributions of the SNe IIn from \cite{2020A&A...637A..73N} along with the SLSN-IIn sample from the PTF are shown in Fig.~\ref{nyholmdist}.

\begin{figure}[h]
   \centering
   \includegraphics[width=8cm]{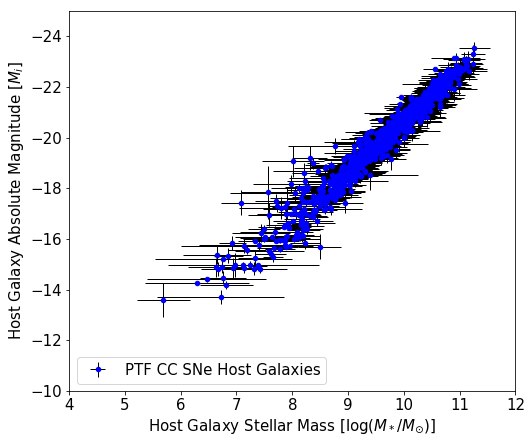}
      \caption{PTF CC SNe host galaxy absolute $i$-band magnitudes are plotted against host galaxy stellar masses. Due to the standard deviation of the residuals being 0.4, which is comparable to the uncertainty on the stellar mass, the $i$-band magnitude can be used as a proxy for the stellar mass in our analysis.} 
         \label{figmimstellar}
   \end{figure}

The BTS is currently the largest spectroscopic survey of SNe. The survey is magnitude-limited in the $g$ and $r$ (<19 magnitude) bands. The sample is 97\% spectroscopically complete at <18 mag, 93\% at <18.5 mag, and 75\% at <19 mag \citep{2020ApJ...904...35P}. The survey is updated daily as new observations come in. For this paper, we chose to use all available CC SNe and IIn from BTS regardless of magnitude as of May 16, 2022. The total number of CC SNe adds up to 949, of which 92 are IIn. We also included the SLSN-II in our analysis, of which there are 18. In BTS, the SLSNe-II are not further divided into subclasses. However, as most SLSNe-II exhibit IIn-like features, we chose to include all of the SLSN-II in our analysis. The parameters we used in our analysis in this paper are redshift, peak magnitude and host $i$-band absolute magnitude. Redshifts are available for all but six CC SNe, none of which are SNe IIn or SLSN-II. Since the host galaxy stellar masses are not available in this sample, we instead utilized the absolute $i$-band magnitude of the hosts where available as these magnitudes are a good proxy for the stellar masses, as is seen in Fig.~\ref{figmimstellar}. Distributions of redshift and host galaxy $i$-band magnitude are shown in Fig.~\ref{btsdist}. 

\begin{figure*}
   \centering
   \includegraphics[width=\textwidth]{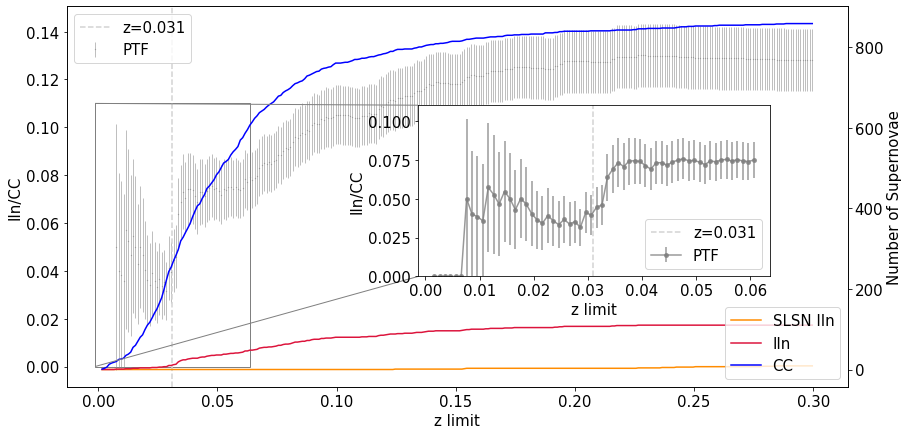}
   \caption{Redshift evolution of SNe-IIn-to-CC ratio. The gray data points represent the cumulative SNe-IIn-to-CC ratio from PTF, indicated on the left y-axis, as a function of redshift limits out to maximum redshift of the sample. SLSNe-IIn are not included in the IIn-to-CC ratio. The vertical gray dashed line represents the chosen redshift cut. The number of CC SNe, SNe IIn, and SLSNe-IIn as a function of redshift are represented by the colored curves indicated on the right y-axis. The inset shows a zoomed-in image of a plot of the cumulative SNe-IIn-to-CC ratio up to redshift 0.06 as well as the chosen redshift cut.}
              \label{Figzlim}%
    \end{figure*}
    
\begin{figure*}
   \centering
   \includegraphics[width=\textwidth]{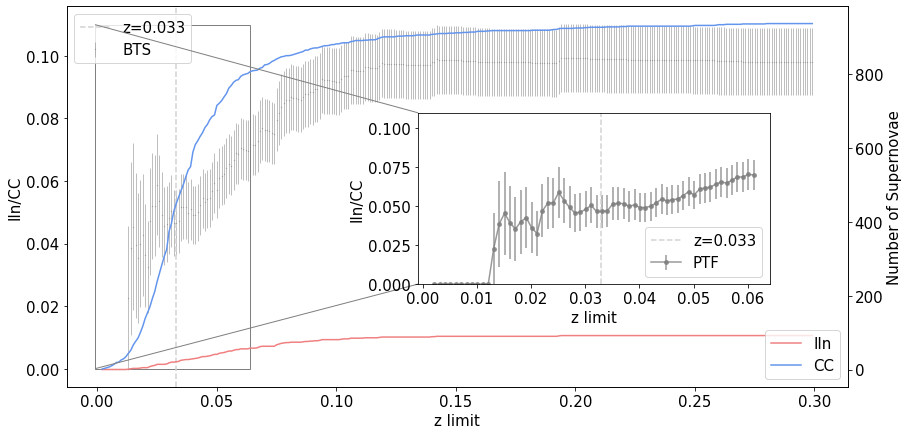}
   \caption{Redshift evolution of SNe-IIn-to-CC ratio. The gray data points represent the cumulative SNe-IIn-to-CC ratio from BTS, indicated on the left y-axis, as a function of redshift limits. The vertical gray dashed line represents the chosen redshift cut. The number of CC SNe and SNe IIn are also depicted as the colored curves, which are indicated on the right y-axis. The inset shows a zoomed-in image of a plot of the cumulative SNe-IIn-to-CC ratio up to redshift 0.06 as well as the chosen redshift cut.}
              \label{Figzlimbts}%
    \end{figure*}

\section{Inferred IIn fractions} \label{section_inferred_fractions}
In \cite{2021MNRAS.500.5142F}, the CC SN rate for the PTF is determined while taking all the survey limitations into account through extensive modeling. This method yields a total of 86 CC SNe and three SNe IIn. Unfortunately, inferring relative rates of SNe IIn with only three sources will be dominated by low-number statistics. For the purpose of estimating the SNe IIn fraction, we will alternatively create an approximately volume-limited sample for the recent PTF data released in \cite{2021ApJS..255...29S} and also from BTS \citep{2020ApJ...895...32F, 2020ApJ...904...35P} under the assumption of fair spectroscopic classification.

A simple way to estimate the distance, or redshift, at which the PTF CC sample is approximately complete is to compare with a known complete sample. The LOSS sample is complete out to 60 Mpc for CC SNe \citep{2011MNRAS.412.1441L, 2017ApJ...837..121G}. With the limiting magnitude of LOSS having a median of $18.8 \pm 0.5$ mag \citep{2011MNRAS.412.1419L}, 60 Mpc is also the distance to the furthest SNe LOSS could theoretically observe assuming the faintest SNe to have an absolute magnitude of around $-15.1$. The PTF has a limiting magnitude of 20.5 mag in the $R$ band, which implies a distance of up to 131.8 Mpc for creating a volume-limited sample, assuming the same $-15.1$ mag for faint SNe. This corresponds to a redshift cut-off of $0.031$ when adopting a Hubble constant of $H_0 = 73 \, \rm{km s^{-1} Mpc^{-1}}$ as in LOSS. This estimate can be tested visually, as is done in Fig.~\ref{Figzlim}, where we plot the SNe-IIn-to-CC ratio as a function of redshift limit. The estimated cut of $0.031$ is indicated in the plot by a dashed line. The SNe IIn fraction for lower redshift cut-offs is dominated by noise due to the small number of supernovae found at these redshifts, whereas the ratio increases above $0.031$ as more SNe IIn are observed at this range. This indicates that the estimated redshift cut is located where the noise from few observations has started to diminish, but we do not yet see the effect of a larger volume wherein to observe SNe IIn and CC SNe in general. As such, imposing a redshift cut-off of $0.031$ on the PTF CC sample is a reasonable approach for creating an approximately volume-limited sample to use for estimating the SNe IIn fraction. 

 \begin{figure}[h]
  \centering
   \includegraphics[width=8cm]{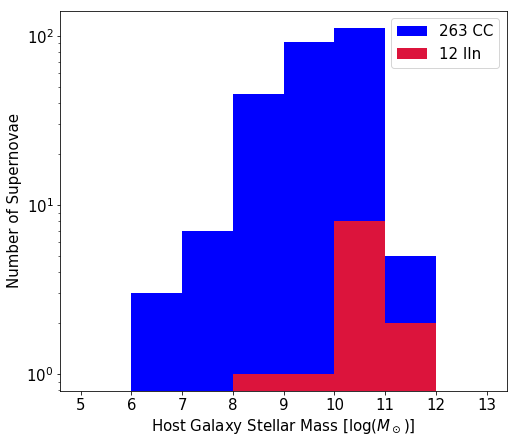}
      \caption{Distributions of host galaxy stellar mass of the complete PTF CC sample using $z < 0.031$. No SLSNe-IIn in the PTF sample are found within this redshift. Results from a KS test show no significant SNe IIn host galaxy mass preference for the volume-limited sample with a p value of 0.002.} 
         \label{fighist}
   \end{figure}

We employ the same method for the BTS sample. Using $-15.1$ mag for the faintest CC SNe is consistent with the mean of the 20 faintest CC SNe in the BTS sample. According to \cite{2019PASP..131a8002B}, the limiting magnitudes for ZTF are 20.8 mag in the $g$ band, 20.6 mag in the $r$ band, and 19.9 in the $i$ band. We will use the $r$-band value to be consistent with LOSS. From the distance modulus, this yields a distance of 138 Mpc, corresponding to a redshift cut-off of about 0.033, as illustrated in Fig.~\ref{Figzlimbts}. The volume effect is less obvious in the BTS data, and as such the resulting IIn fraction from BTS is more robust toward any uncertainties in the redshift cut compared to the result from PTF. However, as the redshift cut of 0.033 occurs before a slow rise in the SNe-IIn-to-CC value and after the inital large uncertainties, we deem 0.033 to be a good estimate for creating an approximately volume-limited sample.   

For the PTF, a redshift cut of 0.031 leaves us with an approximately complete CC sample of 263 CC SNe in total. This includes 12 SNe IIn, but none of the SLSNe-IIn are observed within this redshift. Therefore, we find that the ratio of SNe IIn to CC SNe for our subsample of PTF data to be $0.046 \pm 0.013$. The uncertainty on the resulting SNe-IIn ratio is propagated from the Poisson error on the individual number of CC and SNe IIn. The resulting histogram of the host galaxy stellar masses of this volume limited sample of PTF data, as illustrated in Fig.~\ref{fighist}, reveals no obvious SNe IIn preference for less massive host galaxies, which is in agreement with the analysis done by \cite{2021ApJS..255...29S} on the full PTF sample.  

We can compare this value to the BTS data. A redshift cut-off of 0.033 yields a sample of 440 CC SNe, of which 21 are SNe IIn. This results in an SNe-IIn-to-CC ratio of $0.048 \pm 0.011$. The uncertainty on this number is similarly determined using error propagation. As host galaxy masses are not available in BTS, and a significant amount of CC hosts do not have $i$-band magnitudes either, we will not compare the host galaxy distributions of the SNe IIn and CC SNe from BTS. As these resulting fractions are independent, we combine them and get a SNe IIn relative fraction of $0.047 \pm 0.009$.

\section{Brightness of SNe IIn}
\label{section_brightness}
In this section, we investigate whether the peak brightness of the IIn is influenced by the host galaxy stellar mass when considering the SLSN-IIn as the brighest members of the IIn class. We know from several studies that SLSNe  prefer lower mass host galaxies. According to \cite{2021ApJS..255...29S}, this phenomenon is not significant when only studying the SLSNe-IIn, as the objects are still too few. We note that for redshifts on the order of 0.03 the influence of peculiar velocities on calculating peak absolute magnitudes of the SNe is decreasing compared to SN samples, which are mostly comprised of local sources. 

The distributions of redshift and host galaxy mass of the subsample of IIn and SLSN-IIn from PTF are displayed in Fig.~\ref{nyholmdist}. In Fig.~\ref{btsdist}, we show the distributions of redshift and host galaxy $i$-band magnitude from BTS. For the comparison between these two data sets, we use the $i$-band magnitude of the host galaxies as a proxy for the stellar mass. Only three sources from either data set do not have available host $i$-band magnitudes. 

\begin{figure}[h]
   \centering
   \includegraphics[width=8cm]{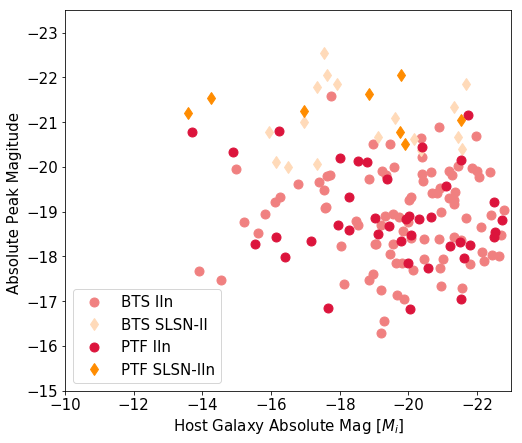}
      \caption{Absolute peak magnitude versus host galaxy $i$-band magnitude for SNe IIn and SLSNe-IIn/SLSNe-II from PTF and BTS.} 
         \label{figmags}
   \end{figure}

\begin{figure}[h]
   \centering
   \includegraphics[width=8cm]{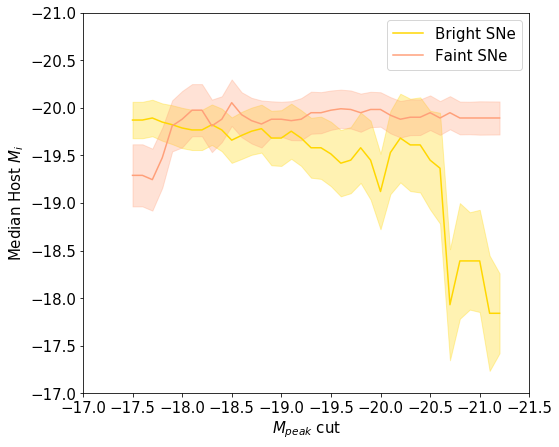}
      \caption{Median host $M_i$ as a function of different cuts on the supernova peak magnitude for combined BTS and PTF samples. The median $i$-band magnitude of the host galaxies becomes fainter for the brighter SNe in the combined sample.  }
         \label{figmedian}
   \end{figure}

In Fig.~\ref{figmags}, we compare the absolute magnitude at peak and the $i$-band magnitude of the host galaxies of the SNe IIn and SLSNe-IIn or SLSN-II from the PTF as well as BTS. As these objects are chosen based on data availability, the subsamples seen in Fig.~\ref{figmags} are not complete. However, \cite{2020A&A...637A..73N} state that the host galaxy mass distribution of their subsample is in agreement with the distribution of the full PTF SNe IIn sample, such that the distribution of the $i$-band magnitudes should follow a similar distribution. The data in Fig.~\ref{figmags} show no clear trend regarding the effect of the host galaxy magnitude on the peak absolute mags of the SNe IIn. To further investigate, we divide the combined PTF and BTS subsamples shown in Fig.~\ref{figmags} in two, namely a faint sample and a bright sample, and subsequently calculate the median and uncertainty on the median as $1.48 \cdot \rm{MAD}/ \sqrt{N-1}$ of the $i$-band magnitudes for the host galaxies, where MAD is the median absolute deviation. This division of the subsample is done for several different SN IIn peak magnitudes. We employ $M_{\mathrm{peak}}$ values from $-17.5$ to $-21$ as the dividing lines between the faint and the bright sub-samples and compare the medians, as can be seen in Fig.~\ref{figmedian}. We find that the median $i$-band magnitude (and thus the stellar mass) of the host galaxies becomes fainter with a $3 \sigma$ significance when choosing a sample of SNe IIn brighter than $-20.5$ mag.  
    
\section{Consequences of a host-mass-dependent IIn fraction}
\label{section_consequences}
While the evidence for a host-mass-dependent IIn fraction is not strong, we next explore the consequences of a hypothetical preference for low-mass host galaxies. We parametrize the IIn fraction as a power-law function of the stellar mass of the host galaxy and investigate the impact on the volumetric rate as a function of redshift. This will be affected since more low-mass galaxies are present in the earlier Universe.

\begin{figure}[h]
   \centering
   \includegraphics[width=8cm]{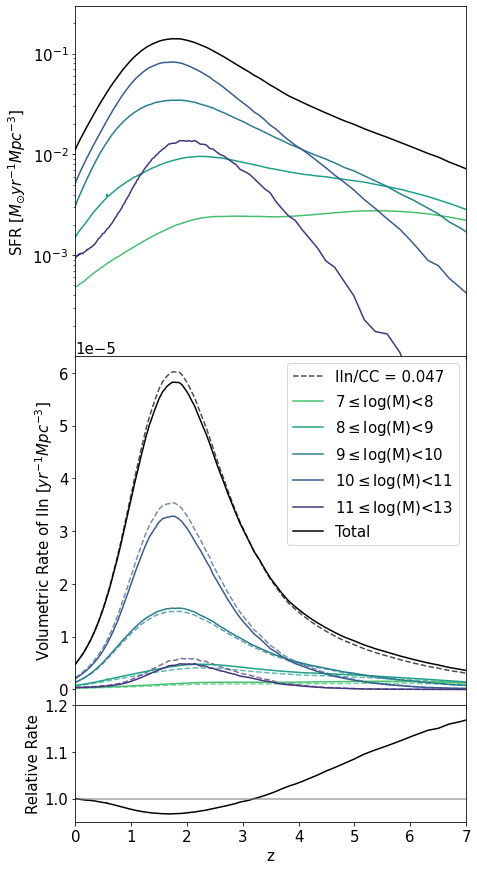}
      \caption{Overview of the cosmic SFR, volumetric SNe IIn rate and relative SNe IIn rate. {\it Top panel}: SFR calculated from \textsc{UniverseMachine} DR1 \citep{2019MNRAS.488.3143B}. {\it Middle panel}: Resulting SNe-IIn rate determined using Eq.~\eqref{eq:volumetric_rate}. The solid lines denote rates from the power-law IIn-to-CC ratio, and the dashed lines represent the rate for a constant ratio of 0.047. The contribution from each mass bin is plotted for comparison. {\it Bottom panel}: Total relative volumetric SNe IIn rate. The gray line represents a line of agreement between the two models.}
         \label{figsfr}
   \end{figure}

In general, the volumetric SNe IIn rate can be expressed as
\begin{equation}
    IIn_{\mathrm{rate}} = \frac{IIn}{CC} \cdot k_{\mathrm{CC}} \cdot SFR.
    \label{eq:volumetric_rate}
\end{equation}
The CC constant, $k_{\mathrm{CC}}$, is set to $0.0091 M_{\odot}^{-1}$ following \cite{2015ApJ...813...93S}. This model is consistent with observational constraints from \cite{2012ApJ...757...70D} and \cite{2014ARA&A..52..415M} as demonstrated in \cite{2015ApJ...813...93S}. Here, a model for the star-formation rate (SFR) is taken from the \textsc{UniverseMachine} algorithm by \cite{2019MNRAS.488.3143B}. The results of this code are best-fitting models of stellar-mass functions (SMFs), cosmic star formation rates (CSFRs), specific star formation rates (sSFRs), and UV luminosity functions (UVLFs) to observations. One can determine the SFR from the output of the \textsc{UniverseMachine} algorithm: 

\begin{equation}
    SFR = \int _{M_{\mathrm{min}}} ^{M_{\mathrm{max}}} SMF \cdot M \cdot sSFR \hspace{5px} \mathrm{d}M .
\end{equation}
When computing the SFR this way, it is possible to split it into different mass bins in order to infer the contribution to the total SFR from galaxies of different masses and how this changes with redshift, as is shown in the top panel of Fig.~\ref{figsfr}. The \textsc{UniverseMachine} resulting models have mass ranges of $10^7 M_\odot$ to $10^{13} M_\odot$, and we choose to split these into five different bins, as indicated in Fig.~\ref{figsfr}. The bin containing the $10^{11} M_\odot$ to $10^{13} M_\odot$ galaxies is chosen to be wider than the other bins, as the contribution from the $10^{12} M_\odot$ to $10^{13} M_\odot$ galaxies is negligible. We parametrize the IIn-to-CC ratio as a power law:
\begin{equation}
    \frac{IIn}{CC} \, (\log(M/M_{\odot})) = 0.15 \cdot \log(M/M_{\odot})^{-0.05} .
\end{equation}
This power-law model is chosen to have a higher SNe-IIn-to-CC ratio than 0.047 for host galaxies below $10^{10} M_{\odot}$ and a lower ratio for more massive galaxies. For this specific example, the ratio will be 0.057 for galaxies with stellar masses of $10^7 M_{\odot}$, and 0.043 for $10^{12} M_\odot$. The motivation for this kind of model comes from the LOSS data in \cite{2017ApJ...837..121G}, showing a preference for low-mass hosts for SNe IIn, which can be modeled with a power law with different sets of parameters \citep{masterthesis}. 

To calculate the volumetric rate, we use the central value of the IIn/CC model for every mass bin in log as the IIn/CC factor in Eq.~\eqref{eq:volumetric_rate}, and thus compute a separate SN IIn rate for each mass bin as well as the combined rate. Since the power-law model and the constant model do not predict the same number of SNe given a different area under the curve, we normalize the resulting rate from the power law model to $4.77 \cdot 10^{-6} \hspace{3pt} \rm{yr^{-1}Mpc^{-3}}$, which is the SNe IIn rate at redshift zero for a constant SNe IIn fraction of 0.047 calculated from Eq.~\eqref{eq:volumetric_rate}, where the \textsc{UniverseMachine} is the source of the SFR. We do this to be consistent with the constant model. The resulting SNe IIn rate is plotted in the bottom panel of Fig.~\ref{figsfr} for both the example power-law model and the constant model of $\rm{IIn/CC} = 0.047$. The overall IIn rate is slightly lower for a redshift below four, and slightly higher for one above four. This plot also shows how the contribution from the different host mass bins differs for the constant ratio to the power-law model. It is evident that the overall rate from low-mass galaxies is higher, and since these contribute a larger fraction of the total rate at higher redshift, we also see an increase in the rate in this high-redshift domain as expected. 

\section{Discussion}
\label{section_discussion}
In this section, we discuss and reflect on some of the shortcomings and consequences of the methods and results of this paper as well as some widely used assumptions. First, we note that the SNe-IIn fractions from the PTF and BTS are based on approximate volume-limited samples. The identification of a sweet spot in the IIn fraction in the PTF sample (Fig.~\ref{Figzlim}) supports this approach, although we note that, ultimately, IIn fractions will have to be based on carefully defined volume-limited samples of a large number of core-collapse supernovae.

From Fig.~\ref{figmedian}, we see a statistically significant connection between the brightness of the SNe IIn and the host galaxy stellar mass when choosing a sample of SNe IIn brighter than $-20.5$ mag. For the faint SNe IIn sample at the $-18$ mag cut-off, the median $i$-band magnitude of the host galaxies drops below the bright IIn sample median host $i$-band magnitude. However, the subsamples employed in this part of our analysis could be subject to different selection effects. Observing faint SNe in bright galaxies is challenging as the light from the galaxy can hide the SNe and so we could expect that some were missed in this area. This effect will influence the median $i$-band host magnitude for the faintest SNe IIn and could explain the slight shift in median host $i$-band magnitude for the faintest SNe in Fig.~\ref{figmedian}. However, the drop in median $i$-band host-galaxy magnitude is not as significant as the drop for the brightest supernovae.  

Using the \textsc{UniverseMachine} algorithm DR1 as the input for the SFR gives insight into the contributions from host galaxies of different stellar masses to the resulting SNe IIn rate whether a preference for low-mass hosts exists or not. One limitation, however, is the lower mass boundary on these galaxies. As is evident from both Figs.~\ref{fighist} and \ref{nyholmdist}, several SNe IIn have host galaxies less massive than $10^{7} M_{\odot}$, but using \textsc{UniverseMachine} it is not possible to see the contribution from these galaxies. Another limitation is the mass resolution of \textsc{UniverseMachine}. In this case, we have four or five data points per mass bin, which prevents us from further dividing the host galaxies into smaller bins to obtain a more detailed overview. 

In Fig.~\ref{figsfr}, we show the resulting SNe-IIn volumetric rates. For illustration, the volumetric rate is computed for a uniform fixed SNe-IIn fraction of 0.047 next to a model in which the SNe IIn prefer lower mass host galaxies, here represented by a power law model. The two models agree at $z=0$ and the normalization of the curves is uncertain by 20 \%. The middle and bottom panels show that the relative volumetric rate of SNe IIn increases over the default constant model for redshifts beyond 3.2.

Introducing the example power-law model to describe the SNe-IIn-to-CC ratio produces a minimal effect on the volumetric rate compared to the constant ratio as seen in Fig.~\ref{figsfr}: A lower rate for redshifts below 3.2 and higher for redshifts beyond. The LSST or \textit{Roman Space Telescope} are not expected to be able to observe lensed SNe beyond redshifts of three and four, respectively \citep{2019MNRAS.487.3342W, 2019ApJS..243....6G, 2021ApJ...908..190P}, and so the possibility of testing such a model is currently limited. On the other hand, we show that a limited mass dependence of the IIn rate should not affect predicted volumetric rates of SNe IIn significantly.

\section{Conclusions}
\label{section_conclusions}
We studied the PTF and BTS SNe IIn and SLSNe-IIn/SLSNe-II populations throughout this work and now present our main conclusions.

Creating a complete sample of CC SNe from PTF and BTS, we find the SNe IIn to CC ratios of $0.046 \pm 0.013$ and $0.048 \pm 0.011$ for the PTF and BTS, respectively. The combined resulting SNe-IIn fraction is $0.047 \pm 0.009$. We see a marginally significant (3 $\sigma$) bias towards low-mass host galaxies for SNe IIn brighter than $-20.5$ mag. We present a general method to evaluate the consequences of a SNe IIn to CC ratio that is nonconstant and is instead described by a power-law model on the resulting volumetric rate. The example model chosen here can be freely replaced by another model as required. We find that the example power law model of $IIn/CC = 0.15 \cdot M^{-0.05}$ results in a slightly lower volumetric rate below a redshift of four and a higher rate beyond a redshift of 3.2 when comparing to a constant ratio of 0.047. Neither the LSST nor the \textit{Roman Space Telescope} are predicted to find lensed SNe beyond redshifts of three or four. We emphasize that our method is generic and can be applied to other CC subtypes if needed.

\begin{acknowledgements}
       We gratefully acknowledge Giorgios Leloudas for sharing peak absolute magnitudes for a subsample of PTF SLSNe-IIn, without which a large part of the analysis in this paper would not have been possible, as well as invaluable comments on the paper draft. We also gratefully acknowledge Steve Schulze and Radek Wojtak for helpful conversations about type IIn and statistical conundrums, respectively, and Wynn Jacobson-Galán and Doogesh Kodi Ramanah for reading and commenting on the paper before submission. We also thank the referee for their useful and thorough comments and suggestions.   
       
       This work was supported by a VILLUM FONDEN Investigator grant to JH (project number 16599).
\end{acknowledgements}

\bibliographystyle{aa} 
\bibliography{biblio} 
\end{document}